\documentclass[10pt]{article}
\usepackage{showkeys}
\begin{document}
\title{{\bf Noncommutativity and Reparametrisation symmetry }}
\author{
{\bf Rabin Banerjee $^{}$\thanks{rabin@bose.res.in}}, {\bf Biswajit Chakraborty $^{}$\thanks{biswajit@bose.res.in}}\\ {\bf Sunandan Gangopadhyay $^{ }$\thanks{sunandan@bose.res.in} }\\
 S.~N.~Bose National Centre for Basic Sciences,\\JD Block, Sector III, Salt Lake, Kolkata-700098, India\\[0.3cm]
}
\date{}

\maketitle

\begin{abstract}
We discuss a general method of revealing both space-space and space-time 
noncommuting structures 
in various models in particle mechanics exhibiting reparametrisation symmetry. 
Starting from the commuting algebra in the conventional gauge, 
it is possible to obtain a noncommuting algebra in a nonstandard gauge. 
The change of variables relating the algebra in the 
two gauges is systematically derived using gauge/reparametrisation 
transformations.
\\[0.3cm]
\\[0.3cm]

\end{abstract}

\section{Introduction}
Issues related to noncommutative space-time in field theories \cite{sz} have led to deep conceptual and technical problems prompting corresponding studies in quantum mechanics. In this context, an important role is played by redefinitions or change of variables which provide a map among the commutative and non-commutative structures \cite{np},\cite{dh},\cite{pin},\cite{rb}. However, there does not seem to be a precise underlying principle on which such maps are based. One of the motives of this paper is to provide a systematic formulation of such maps. In the models discussed here, these maps are essentially gauge/reparametrisation transformations.

A general feature indicated by this analysis is the possibility of noncommuting space-space (or space-time) coordinates for models in particle mechanics with 
reparametrisation symmetry. The point to note is that even if the model does not have this symmetry naturally, it can always be introduced by hand as, for instance, in the nonrelativistic (NR) free particle. We shall discuss this example in details and reveal the various noncommuting structures. As other examples, we consider the free relativistic particle as well as its interaction with a background electromagnetic field.

We exploit the reparametrisation invariance to find a nonstandard gauge in which the space-time and/or space-space coordinates become noncommuting. In contrast to recent approaches \cite{pin}, we provide a definite method of finding this gauge. We also show that the variable redefinition relating the nonstandard and standard gauges is a gauge transformation.

In section 2, we discuss how any particle model can be rewritten in a time reparametrisation invariant form. This is used to show the occurence of noncommuting structures in the usual nonrelativistic free particle model. The free relativistic particle is analysed in section 3. Here we also analyse
 the structure of the angular momentum operator in some details.
 A gauge independent expression is obtained, which therefore does not require
any central extension in the non-standard gauge. The interaction of the
 free relativistic particle with an external electromagnetic field
 is considered in section 4. Finally, we conclude in section 5. 

There are two appendices. In appendix A, we establish the connection between
Dirac brackets in the axial and radiation gauges using suitable gauge
transformations. In appendix B, we show, in the symplectic formalism,
 the connection between integral curves and the equations of motion in the time
reparametrised version. This also shows how constraints come into picture
in the time reparametrised formulation.

\section{Particle models}
Consider the action for a point particle in classical mechanics
\begin{equation}
S[x(t)] = \int_{t_{1}}^{t_{2}} dt L\left(x, \frac{dx}{dt}\right)
\label{Z1}
\end{equation}
The above form of the action can be rewritten in a time-reparametrised invariant
form by elevating the status of time $t$ to that of an additional variable,
along with $x$, in the configuration space as,
\begin{equation}
S[x(\tau), t(\tau) ] = \int_{\tau_{1}}^{\tau_{2}} d\tau \dot{t} L\left(x, \frac{\dot{x}}{\dot{t}}\right) = \int_{\tau_{1}}^{\tau_{2}} d\tau L_{\tau}\left(x, \dot{x}, t, \dot{t}\right)
\label{Z2}
\end{equation}
where,
\begin{equation}
L_{\tau}(x, \dot{x}, t, \dot{t}) = \dot{t}L\left(x, \frac{\dot{x}}{\dot{t}}\right)  \quad; \quad \dot{x} = \frac{dx}{d\tau} \quad \dot{t} = \frac{dt}{d\tau}  
\label{Z3}
\end{equation}
and $\tau$ is the new evolution parameter that can be taken to be an arbitrary monotonically increasing function of time $t$.
Now the canonical momenta corresponding to the coordinates $t$ and $x$ are given by
\begin{eqnarray}
p_{t} &=& \frac{\partial L_{\tau}}{\partial \dot{t}} =  L\left(x, \frac{\dot x}{\dot t}\right) + \dot{t}\frac{\partial L\left(x, \frac{\dot x}{\dot t}\right)}{\partial \dot t}\nonumber\\
&=& L\left(x, \frac{dx}{dt}\right) - \frac{dx}{dt}\frac{\partial L(x, dx/dt)}{\partial (dx/dt)} = -H
\label{Z4}
\end{eqnarray}
\begin{equation}
p_{x} = \frac{\partial L_{\tau}}{\partial \dot{x}} 
\label{Z5}
\end{equation}
As happens for a time-reparametrised theory, the canonical Hamiltonian (using (\ref{Z4}, \ref{Z5})) vanishes:
\begin{equation}
H_{\tau} = p_{t}\dot{t} + p_{x}\dot{x} - L_{\tau} = \dot{t}(H + p_{t}) = 0.
\label{N6}
\end{equation}

As a particular case of (\ref{Z1}), we start from the action of a free NR particle in one dimension
\begin{equation}
S = \int dt \frac{1}{2}m\left(\frac{dx}{dt}\right)^{2}
\label{N1}
\end{equation}
The above form of the action can be rewritten in a time-reparametrised invariant
form as in (\ref{Z2}), 
\begin{equation}
S = \int d\tau L_{\tau}(x, \dot{x}, t, \dot{t})
\label{N2}
\end{equation}
where
\begin{equation}
L_{\tau}(x, \dot{x}, t, \dot{t}) = \frac{m}{2}\frac{\dot{x}^2}{\dot{t}}\qquad; \qquad \dot{x} = \frac{dx}{d\tau} \quad \dot{t} = \frac{dt}{d\tau}  
\label{N3}
\end{equation}
and $\tau$ is the new evolution parameter that can be taken to be an arbitrary monotonically increasing function of time $t$.
Now the canonical momenta corresponding to the coordinates $t$ and $x$ are given by
\begin{equation}
p_{t} = \frac{\partial L_{\tau}}{\partial \dot{t}} = -\frac{m\dot{x}^2}{2\dot{t}^2}
\label{N4}
\end{equation}
\begin{equation}
p_{x} = \frac{\partial L_{\tau}}{\partial \dot{x}} = \frac{m\dot{x}}{\dot{t}}
\label{N5}
\end{equation}
which satisfy the standard canonical Poisson bracket (PB) relations
\begin{equation}
\{x, x\} = \{p_{x}, p_{x}\} = \{t, t\} = \{p_{t}, p_{t}\} = 0 \quad \{x,p_{x}\} = \{t, p_{t}\} = 1
\label{N500}
\end{equation}

As happens for a time-reparametrised theory, the canonical Hamiltonian (using (\ref{N4}, \ref{N5})) vanishes:
\begin{equation}
H_{\tau} = p_{t}\dot{t} + p_{x}\dot{x} - L_{\tau} = 0.
\label{N6}
\end{equation}
 Also, the primary constraint in the theory, obtained from (\ref{N4}, \ref{N5}) is given by
\begin{equation}
\phi_{1} = p_{x}^2 + 2mp_{t} \approx 0
\label{N7}
\end{equation}
where $\approx 0$ implies equality in the weak sense \cite{dirac}, \cite{dir}. Clearly the space-time
coordinate $x^{\mu}(\tau)$, ($\mu = 0, 1 ; x^0 = t, x^1 = x$), transforms as
a scalar under reparametrisation:
\begin{eqnarray}
\tau \rightarrow \tau^{'} &=& \tau^{'}(\tau)\nonumber\\
x^{\mu}(\tau)\rightarrow x^{'\mu}(\tau^{'}) &=& x^{\mu}(\tau)
\label{N71}
\end{eqnarray}
Consequently under an infinitesimal reparametrisation transformation ($\tau^{'} = \tau - \epsilon$), the infinitesimal change in the space-time coordinate is given by
\begin{eqnarray}
\delta x^{\mu}(\tau) = x^{'\mu}(\tau) - x^{\mu}(\tau) = \epsilon \frac{dx^\mu}{d\tau} 
\label{N8}
\end{eqnarray}
The generator of this reparametrisation transformation is obtained by first writing the variation in the Lagrangian $L_\tau$ (\ref{N3}) under the transformation (\ref{N8}) as a total derivative,
\begin{eqnarray}
\delta L_\tau = \frac{dB}{d\tau} \qquad; \qquad B = \frac{m\epsilon}{2} \frac{\dot x^2}{\dot t}
\label{N11}
\end{eqnarray}
Now the generator $G$ is obtained from the usual Noether's prescription as,
\begin{eqnarray}
G = p_{t}\delta t + p_{x}\delta x - B = \frac{\epsilon \dot t}{2m}\phi_{1}
\label{N12}
\end{eqnarray}
It is easy to see that this generator reproduces the appropriate transformation (\ref{N8})
\begin{eqnarray}
\delta x^{\mu}(\tau) = \{x^{\mu}, G\} = \epsilon \frac{dx^\mu}{d\tau} 
\label{N12a}
\end{eqnarray}
which is in agreement with Dirac's treatment \cite{dirac}, \cite{dir}\footnote{In this treatment, the generator is a linear combination of the first class constraints. Since we have only one first class constraint $\phi_{1}$, the gauge generator is proportional to $\phi_{1}$}. Note that $x^{\mu}$'s are not gauge invariant variables in this case. This
example shows that reparametrisation symmetry can be identified with gauge symmetry.\\
Let us now fix the gauge symmetry by imposing a gauge condition. The standard choice is to identify the time coordinate $t$ with the parameter $\tau$, 
\begin{equation}
\phi_{2} = t - \tau \approx 0.
\label{N13}
\end{equation}
The constraints (\ref{N7}, \ref{N13}) form a second class set with
\begin{equation}
\phi_{ab} = \{\phi_{a}, \phi_{b}\} = -2m\epsilon_{ab}\quad (a, b = 1, 2)
\label{N14}
\end{equation}
where, $\epsilon_{ab}$ is an anti-symmetric tensor with $\epsilon_{12} = 1$.\\
The next step is to compute the Dirac brackets (DB) defined as,
\begin{equation}
\{A, B\}_{DB} = \{A, B\} - \{A, \phi_{a}\}(\phi^{-1})_{ab}\{\phi_{b}, B\} 
\label{N15}
\end{equation}
where $A$, $B$ are any pair of phase-space variables and $(\phi^{-1})_{ab} = (2m)^{-1}\epsilon_{ab}$ is the inverse of $\phi_{ab}$. It then follows,
\begin{equation}
\{x, x\}_{DB} = \{p_{x}, p_{x}\}_{DB} = 0 \qquad \{x, p_{x}\}_{DB} = 1
\label{N16}
\end{equation}
This reproduces the expected canonical bracket structure in the usual $2-d$ reduced phase-space comprising of variables $x$ and $p_{x}$ only. The DB imply a strong imposition of the second class constraints ($\phi_{a}$). Consistent with this, $\{t, x\}_{DB} = 0$ showing that there is no space-time non-commutativity if a gauge-fixing condition like (\ref{N13}) is chosen. A natural question arises is whether space-time (or space-space) non-commutativity can be obtained by imposing a suitable variant of the gauge fixing condition (\ref{N13}). 
Before answering this question, we emphasize that the DB between various gauges should be related by suitable gauge transformations\footnote{We show (see appendix A) how this is done for a free Maxwell theory where the DB between phase-space variables in radiation and axial gauges are related by appropriate gauge transformations.}. This idea will be useful.

 In the present case, to get hold of a set of variables $x^{'}$, $t^{'}$ satisfying a noncommutative algebra,
\begin{eqnarray}
\{t^{'}, x^{'}\}_{DB} = \theta
\label{N17}
\end{eqnarray}
with $\theta$ being constant, the same procedure, as done (in the appendix) for a free Maxwell theory, is adopted. The transformations (\ref{N8}) are written in terms of phase-space variables, after strongly implementing the constraint (\ref{N13}). Then, in component notation,
\begin{eqnarray}
t^{'} = t + \epsilon
\label{N18}
\end{eqnarray}
\begin{eqnarray}
x^{'} = x + \epsilon\frac{dx}{d\tau} = x + \epsilon\frac{p_{x}}{m}.
\label{N19}
\end{eqnarray}
Substituting back in the L.H.S. of (\ref{N17}) and using the Dirac algebra (\ref{N16}) for the unprimed variables, fixes $\epsilon$ as,
\begin{eqnarray}
\epsilon = -\theta p_{x}.
\label{N20}
\end{eqnarray}
 This shows that the desired gauge fixing condition is 
\begin{eqnarray}
t^{'} + \theta p_{x} - \tau \approx 0
\label{N21}
\end{eqnarray}
 Now one can just drop the prime to rewrite (\ref{N21}) as 
\begin{eqnarray}
t + \theta p_{x} - \tau \approx 0
\label{N2111}
\end{eqnarray}
 Expectedly, a direct calculation of the Dirac bracket in this gauge immediately reproduces the noncommutative structure $\{t, x\}_{DB} = \theta$.

 This analysis can be generalised trivially to higher $d + 1$-dimensional Galilean space-time. In the case of $d \geq 2$, one can see that the above space-time noncommutativity is of the form $\{x^{0}, x^i\}_{DB} = \theta^{0i}$; ($x^{0} = t$). This can be derived by writing the transformations (\ref{N18}, \ref{N19}) for $d \geq 2$ as,
\begin{eqnarray}
x^{'0} = x^{0} + \epsilon
\label{N18a}
\end{eqnarray}
\begin{eqnarray}
x^{'i} = x^{i} + \epsilon\frac{dx^{i}}{d\tau} = x^{i} + \epsilon\frac{p^{i}}{m}.
\label{N19b}
\end{eqnarray}
which, when substituted back in the L.H.S. of $\{x^{'0}, x^{'i}\} = \theta^{0i}$, fixes $\epsilon$ as
\begin{eqnarray}
\epsilon = -\theta^{0i}p_{i}.
\label{N19c}
\end{eqnarray}
The desired gauge fixing condition (dropping the prime) now becomes
\begin{eqnarray}
x^{0} + \theta^{0i}p_{i} - \tau \approx 0
\label{N19d}
\end{eqnarray}
which is the analogue of (\ref{N2111}). For $d \geq 2$, the space-space algebra is also NC 
\begin{eqnarray}
\{x^{i}, x^{j}\}_{DB} = -\frac{1}{m}\left(\theta^{0i}p^{j}- \theta^{0j}p^{i} \right).
\label{N211}
\end{eqnarray}
The remaining non-vanishing DB(s) are
 \begin{eqnarray}
\{x^{i}, p_{0}\}_{DB} = -\frac{p^{i}}{m}\qquad \{x^{i}, p_{j}\}_{DB} = {\delta^{i}}_{j}.
\label{N211a}
\end{eqnarray}
The above structures of the Dirac brackets show a Lie-algebraic structure
 for the
brackets involving phase-space variables (with the inclusion of identity). Following \cite{madore}, one can
therefore associate an appropriate ``diamond product" for this, in order to
compose any pair of phase-space functions.

We have thus systematically derived the nonstandard gauge condition leading to a noncommutative algebra. Also, the change of variables mapping this noncommutative algebra with the usual (commutative) algebra is found to be a gauge transformation.

There is another interesting way of deriving the Dirac algebra if one looks
at the symplectic two form $\omega = dp_{\mu}\wedge dx^{\mu}$ and then simply
impose the conditions on $p_{0}$ and $x^{0}$, for all cases discussed. We
consider the simplest case here. In $1+1$-dimension, the two form $\omega$ can
be written as
\begin{eqnarray}
\omega = dp_{t}\wedge dt +  dp_{x}\wedge dx
\label{N211b}
\end{eqnarray}
Now imposing the condition on $p_{t}$ (\ref{N7}) and $t$ (\ref{N13}), we get,
\begin{eqnarray}
\omega = -\frac{p_{x}}{m}dp_{x}\wedge d\tau + dp_{x}\wedge dx  
\label{N211c}
\end{eqnarray}
Note that the first term on the right hand side of the above equation vanishes
as $\tau$ is not a variable in the configuration space. Now the inverse of the
components of the two form yields the brackets (\ref{N16}). 

Next we carry out the above analysis in the non-standard gauge (\ref{N2111}). In
this case, after imposing the condition on $p_{x}$ from (\ref{N2111}),
 the two form $\omega$ reads,
\begin{eqnarray}
\omega = dp_{t}\wedge dt - \frac{1}{\theta} dt\wedge dx  
\label{N211d}
\end{eqnarray}
Once again, a straight forward computation of the inverse of the
components of the two form yields the  noncommutative structure
 $\{t, x\} = \theta$.
The same procedure can be followed for the other cases
discussed in the paper.

The role of integral curves within this symplectic formalism \cite{sud} 
is discussed in appendix B.

\section{Relativistic Free Particle}
In this section we take up the case of a free relativistic particle and study how space-time noncommutativity can arise in this case also through a suitably modified gauge fixing condition. To that end, we start with the standard reparametrisation invariant action of a relativistic free particle which propagates in $d + 1$-dimensional ``target
 spacetime" 
\begin{equation}
S_{0} = -m\int d\tau \sqrt{-\dot{x}^{2}}
\label{1}
\end{equation}
with space-time coordinates  $x^{\mu}$, ${\mu} = 0, 1, ...d$, the dot denoting
differentiation with respect to the evolution parameter $\tau$, and the Minkowski metric is $\eta = diag(-1, 1, ..., 1)$. Note that here it is already in the reparametrised form with all $x^{\mu}$'s (including $x^{0} = t$) contained in the configuration space. The canonically conjugate momenta are given by
\begin{equation}
p_{\mu} = \frac{m\dot{x}_{\mu}}{\sqrt{-\dot{x}^{2}}}
\label{2}
\end{equation}
and satisfy the standard PB relations
\begin{equation}
\{x^{\mu}, p_{\nu}\} = \delta^{\mu}_{\nu}  \qquad \{x^{\mu}, x^{\nu}\} = \{p^{\mu}, p^{\nu}\} = 0.
\label{3}
\end{equation}
These are subject to the Einstein constraint
\begin{equation}
\phi_{1} = p^{2} + m^{2} \approx 0
\label{4}
\end{equation}
which follows by taking the square of (\ref{2}).
Now using the reparametrisation symmetry of the problem (under which the action (\ref{1}) is invariant) and the fact that $x^{\mu}(\tau)$ transforms as a  scalar under world-line reparametrisation (\ref{N71}), again leads to the infinitesimal transformation of the space-time coordinate (\ref{N8}).
As before, to derive the generator of the reparametrisation invariance 
we write the variation in the Lagrangian as a total derivative,  
\begin{eqnarray}
\delta L &=& \frac{dB}{d\tau}\qquad;\qquad B = -m\epsilon\sqrt{-\dot{x}^2}
\label{4d}
\end{eqnarray}
The generator is obtained from the usual Noether's prescription\footnote {The factor of $1/2$ comes from symmetrisation. To make this point clear, we must note that while computing $\{x^{\mu}, G\}$, an additional factor of 2 crops up from the bracket between $x^{\mu}$ and $\delta x_{\mu}$ as $\delta x_{\mu}$ is related to $p_{\mu}$ by the relations (\ref{N8}) and (\ref{2}). The factor of $1/2$ is placed in order to cancel this additional factor of $2$.}
\begin{eqnarray}
G &=& \frac{1}{2}\left(p^{\mu}\delta x_{\mu} - B\right) = \frac{\epsilon\sqrt{-\dot{x}^2}}{2m}\phi_{1}
\label{5a}
\end{eqnarray}
where we have used (\ref{N8}, \ref{4d}). Clearly we find that $G$ generates
the infinitesimal transformation of the space-time coordinate (\ref{N12a}).
Now we can impose a gauge condition to curtail the gauge freedom just as in the NR case. The standard choice is to identify the time coordinate $x^{0}$ with the parameter $\tau$, 
\begin{equation}
\phi_{2} = x^{0} - \tau \approx 0
\label{5}
\end{equation}
which is the analogue of (\ref{N13}).
The constraints (\ref{4}, \ref{5}) form a second class set with
\begin{equation}
\{\phi_{a}, \phi_{b}\} = 2p_{0}\epsilon_{ab}
\label{6}
\end{equation}
The resulting non-vanishing DB(s) are
\begin{eqnarray}
\{x^{i}, p_{0}\}_{DB} = \frac{p^{i}}{p_{0}} \qquad\{x^{i}, p_{j}\}_{DB} = {\delta^{i}}_{j}
\label{8000}
\end{eqnarray}
which imposes the constraints $\phi_{1}$ and $\phi_{2}$ strongly. In particular,  we observe $\{x^0, x^i\}_{DB} = 0$, showing that there is no space-time noncommutativity. This is again consistent with the fact that the constraint (\ref{5}) is now strongly imposed. Taking a cue from our previous NR example, we see that we must have a variant of (\ref{5}) as a gauge fixing condition to get space-time noncommutativity in the following form
\begin{eqnarray}
\{x^{'0}, x^{'i}\}_{DB} = \theta^{0i}
\label{9a}
\end{eqnarray}
($\theta^{0i}$ being constants) where $x^{'\mu}$ denotes the appropriate gauge
transforms of $x^{\mu}$ variables. To determine these transformed variables $x^{'\mu}$ in terms of the variables  $x^{\mu}$, we consider an
infinitesimal transformation (\ref{N8}) written in terms of phase-space variables as 
\begin{eqnarray}
x^{'0} = x^{0} + \epsilon \qquad;\qquad x^{'i} &=& x^{i} - \epsilon\frac{p^{i}}{p_{0}}
\label{9b}
\end{eqnarray}
where we have used the relation $\frac{dx^{i}}{d\tau} = -\frac{p^{i}}{p_{0}}$
obtained from (\ref{2}).
Substituting the above relations (\ref{9b}) back in (\ref{9a}) and using (\ref{8000}), a simple inspection shows that $\epsilon$ is given by
\begin{eqnarray}
\epsilon = -\theta^{0i}p_{i}
\label{9d}
\end{eqnarray}
which is identical to (\ref{N19c}).
Hence the gauge transformed variables $x^{'\mu}$ (\ref{9b}) for the above choice of $\epsilon$ are given by
\begin{eqnarray}
x^{'0} = x^{0} - \theta^{0i}p_{i}
\label{10e}
\end{eqnarray}
\begin{eqnarray}
x^{'i} &=& x^{i} + \theta^{0j}p_{j}\frac{p^{i}}{p_{0}}
\label{10f}
\end{eqnarray}
Using the above set of transformations and the relation (\ref{8000}), we obtain
the Dirac algebra between the primed variables,
\begin{eqnarray}
\{x^{'0}, x^{'i}\}_{DB} = \theta^{0i} 
\label{10a}
\end{eqnarray}
\begin{eqnarray}
\{x^{'i}, x^{'j}\}_{DB} = \frac{1}{p_{0}}\left(\theta^{0i}p^{j} - \theta^{0j}p^{i}\right) 
\label{10b}
\end{eqnarray}
\begin{eqnarray}
\{x^{'i}, p^{'}_{0}\}_{DB} = \frac{p^{i}}{p_{0}}\qquad; \qquad\{x^{'i}, p^{'}_{j}\}_{DB} = {\delta^{i}}_{j} 
\label{10c}
\end{eqnarray}
Note that unlike $x$'s, $p$'s are gauge invariant objects as $\{p^{\mu}, \phi\} = 0$; hence $p^{'}_{\mu} = p_{\mu}$.

It is interesting to observe that the solution of the gauge parameter $\epsilon$ remains the same in both the relativistic case as well as the NR case. Also, $m$ in the NR case gets replaced by $-p_{0}$ in the relativistic case. With this identification, one can easily see that the complete Dirac algebra in the NR case goes over  to the corresponding algebra in the relativistic case. However,
since $p_{0}$ does not have a vanishing bracket with all other phase-space
variables, its occurence in the denominators in (\ref{10b}, \ref{10c})
 shows that the bracket structure of the phase-space variables in the
relativistic case is no longer Lie-algebraic, unlike the NR case discussed
in the previous section.

Furthermore, the modified gauge fixing condition is given by
\begin{eqnarray}
\phi_{2} = x^{0} + \theta^{0i}p_{i} - \tau \approx 0, \qquad i = 1, 2, ...d
\label{9e}
\end{eqnarray}
It is trivial to check that the constraints (\ref{4}, \ref{9e}) also form
a second class pair as
\begin{eqnarray}
\{\phi_{a}, \phi_{b}\} = 2p_{0}\epsilon_{ab}. 
\label{10}
\end{eqnarray}
The set of non-vanishing DB(s) consistent with the strong imposition of the constraints (\ref{4}, \ref{9e}) reproduces the results (\ref{10a}, \ref{10b}, \ref{10c}).
(\ref{10c}) is the same as in the standard gauge (\ref{5}), while (\ref{10b}) implies non-trivial commutation relations among spatial coordinates upon quantisation. 

It should be noted that the above gauge fixing condition (\ref{9e}) was also given in \cite{pin}. Indeed a change of variables, which is different from (\ref{10e}, \ref{10f}), is found there by inspection, using which the space-time  noncommutativity gets removed. However, the change of variables given in this paper is related to a gauge transformation which in turn gives a systematic derivation of the modified gauge condition and also space-time noncommutativity. Moreover, their \cite{pin} definition of the Lorentz generators (rotations and boosts) requires some additional terms (in the modified gauge) in order to have a closed algebra between the generators. In our approach, the definition of the Lorentz generators remains unchanged, simply because these are gauge invariant.

 The Lorentz generators (rotations and boosts) are defined as,
\begin{eqnarray}
M_{ij} = x_{i}p_{j} - x_{j}p_{i} 
\label{27}
\end{eqnarray}
\begin{eqnarray}
M_{0i} = x_{0}p_{i} - x_{i}p_{0} 
\label{28}
\end{eqnarray}
Expectedly, they satisfy the usual algebra in both the unprimed and the primed coordinates as $M_{\mu\nu}$ and $p_{\mu}$ are both gauge invariant.
\begin{eqnarray}
\{M_{ij}, p_{k}\}_{DB} = \delta_{ik}p_{j} - \delta_{jk}p_{i} 
\label{29}
\end{eqnarray}
\begin{eqnarray}
\{M_{ij}, M_{kl}\}_{DB} = \delta_{ik}M_{jl} - \delta_{jk}M_{il} + \delta_{jl}M_{ik} - \delta_{il}M_{jk}
\label{29a}
\end{eqnarray}
\begin{eqnarray}
\{M_{ij}, M_{0k}\}_{DB} = \delta_{ik}M_{0j} - \delta_{jk}M_{0i} 
\label{29b}
\end{eqnarray}
\begin{eqnarray}
\{M_{0i}, M_{0j}\}_{DB} = M_{ji} 
\label{29c}
\end{eqnarray}
However, the algebra between the space coordinates and the rotations, boosts
are different in the two gauges (\ref{5}, \ref{9e}). This is expected as $x^{k}$ is not gauge invariant under gauge transformation. We find,
\begin{eqnarray}
\{M_{ij}, x^{k}\}_{DB} = {\delta_{i}}^{k}x_{j} - {\delta_{j}}^{k}x_{i} 
\label{30}
\end{eqnarray}
\begin{eqnarray}
\{M_{0i}, x^{j}\}_{DB} = x_{i}\frac{p^{j}}{p_{0}} - x_{0}{\delta_{i}}^{j} 
\label{31}
\end{eqnarray}
\begin{eqnarray}
\{M_{ij}, x^{'k}\}_{DB} &=& \{M_{ij}, x^{k} + \theta^{0l}p_{l}\frac{p^{k}}{p_{0}}\}_{DB}\nonumber\\
&=& {\delta_{i}}^{k}x^{'}_{j} - {\delta_{j}}^{k}x^{'}_{i} + \frac{1}{p_{0}}\left({\theta^{0}}_{i}p^{k}p_{j} - {\theta^{0}}_{j}p^{k}p_{i}\right) 
\label{32}
\end{eqnarray}
\begin{eqnarray}
\{M_{0i}, x^{'j}\}_{DB} &=& \{M_{0i}, x^{j} + \theta^{0l}p_{l}\frac{p^{j}}{p_{0}}\}_{DB}\nonumber\\
&=& x^{'}_{i}\frac{p^{j}}{p_{0}} - x^{'}_{0}{\delta_{i}}^{j} - {\theta^{0}}_{i}p^{j} 
\label{33}
\end{eqnarray}
where we have used the algebra $(\ref{8000})$ followed by (\ref{10f}).
The same results can also be obtained using the relations (\ref{10a}, \ref{10b},\ref{10c}).

The gauge choice (\ref{9e}) is not Lorentz invariant.
Yet the Dirac bracket procedure forces this constraint equation to be strongly valid in all Lorentz frames \cite{dir}. This can be made consistent if and only if an infinitesimal Lorentz boost to a new frame \footnote{A similar treatment has been given in \cite{bc} for a free relativistic particle coupled to Chern-Simons term.} 
\begin{eqnarray}
p^{\mu} \rightarrow p^{'\mu} = p^{\mu} + \omega^{\mu\nu}p_{\nu}
\label{34}
\end{eqnarray}
is accompanied by a compensating infinitesimal gauge transformation
\begin{eqnarray}
\tau \rightarrow \tau^{'} = \tau + \Delta\tau
\label{35}
\end{eqnarray}
The change in $x^{\mu}$, upto first order, is therefore
\begin{eqnarray}
x^{'\mu}(\tau) & = &x^{\mu}(\tau^{'}) + \omega^{\mu\nu}x_{\nu}(\tau)\nonumber\\
 & = & x^{\mu}(\tau) + \Delta\tau\frac{dx^{\mu}}{d\tau} + \omega^{\mu\nu}x_{\nu}
\label{36}
\end{eqnarray}
In particular, the zero-th component is given by,
\begin{eqnarray}
x^{'0}(\tau) = x^{0}(\tau) + \Delta\tau\frac{dx^{0}}{d\tau} + \omega^{0i}x_{i}
\label{37}
\end{eqnarray}
Since the gauge condition (\ref{9e}) is $x^{0}(\tau) \approx \tau - \theta^{0i}p_{i}$, $x^{'0}(\tau)$ also must satisfy $x^{'0}(\tau) = (\tau - \theta^{0i}p^{'}_{i})$ in the boosted frame, which can now be written, using (\ref{34}), as
\begin{eqnarray}
x^{'0}(\tau) &=& \tau - \theta^{0i}p^{'}_{i}\nonumber\\
 &=& \tau - \theta^{0i}p_{i} + \theta^{0i}\omega^{0i}p_{0}.
\label{38}
\end{eqnarray}
Comparing with (\ref{37}) and using the gauge condition (\ref{9e}), we can now solve for $\Delta \tau$ to get,
\begin{eqnarray}
\Delta\tau = \frac{\theta^{0i}\omega^{0i}p_{0} - \omega^{0i}x_{i}}{1 - \theta^{0i}\dot{p_{i}}}\qquad; \quad \dot p_{i} = \frac{dp_{i}}{d\tau}
\label{39}
\end{eqnarray}
Therefore, for a pure boost, the spatial components of (\ref{36}) satisfy
\begin{eqnarray}
\delta x^{j}(\tau) &=& x^{'j}(\tau) - x^{j}(\tau) = \Delta\tau\frac{dx^{j}}{d\tau} + \omega^{j0}x_{0}\nonumber\\
&=& \omega^{0i}\left(x_{i}\frac{p^{j}}{p_{0}} - x_{0}{\delta_{i}}^{j} - \theta^{0i}p^{j}\right)
\label{40}
\end{eqnarray}
Hence we find that (\ref{40}) and (\ref{33}) are consistent with each other. However, note that in the above derivation we have taken $\theta^{0i}$ to be a constant. If we take $\theta^{0i}$ to transform as a tensor, then for a Lorentz boost to a new frame, it changes as,
\begin{eqnarray}
\theta^{0i} \rightarrow \theta^{'0i} = \theta^{0i} + \omega^{0j}\theta^{ji} 
\label{41}
\end{eqnarray}
and the entire consistency program would fail. The ($1 + 1$)- dimensional case is special since even if we take $\theta^{01}$ to transform as a tensor, this will not affect the consistency program as it remains invariant ($\theta^{'01} = \theta^{01}$) under Lorentz boost.

Let us now make certain observations.
 Although, the relations (\ref{N211}) and (\ref{10b}) are reminescent of Snyder's algebra \cite{snyder}, there is a subtle
difference. To see this, note that the right hand side of these relations
 do not have the structure of an angular momentum
 operator in their differential representation
 (obtained by repacing $p_{j}$ by ($-i\partial_{j}$) in contrast to the Snyder's
algebra. 

 Now in the cases where the noncommutativity takes the canonical structure
 ($[\hat x^\mu,\hat x^\nu]=i\theta^{\mu\nu}$), the presence of non-locality
is inferred from the fact that two  localised functions
 $f$ and $g$ having supports within a size
 $\delta << \sqrt{||\theta||}$, yields a function $f\star g$
 which is non-vanishing over a much larger region of size $||\theta||/\delta$ 
\cite{sz}. One therefore expects a similar qualitative feature 
of non-locality arising from the ``diamond product" appropriate for the
Lie bracket structure of noncommutativity in the NR case also.
 This is further reinforced by the fact
that coordinate transformations (\ref{N18a}, \ref{N19b}) involve mixing of
coordinates and momenta. Since this mixing is present in the relativistic case
as well (\ref{10e}, \ref{10f}), it is expected to maintain the non-locality of the
noncommutative theory, although an appropriate ``diamond product" cannot be readily 
constructed because of the absence of a Lie bracket structure. Also, the mixing
of coordinates and momenta is a natural consequence of our gauge conditions
which essentially involve phase-space variables interpolating between the
commutative and noncommutative descriptions.
Note however, the transformed coordinates (\ref{N18a}, \ref{N19b}) are distinct from the covariant coordinates $\hat X^i=\hat x^i + \theta^{ij}\hat A_{j}$, (where $\hat A_{j}$
 is a noncommutative gauge field)
introduced in \cite{madore}, at the noncommutative field theoretical level, to render the transformation property of the
 product $\hat X^i \psi$ covariant just like the field $\psi(\hat x^i)$. This is because $\hat A_{i}$ cannot be identified with $\hat p_{\mu}$, as at the noncommutative field theoretical level
 one does not have any $\hat p_{\mu}$ conjugate to $\hat x^\mu$ since $\hat x^\mu$'s are just
set of operator valued $q$-numbers labelling the  
degrees of freedom in the system and are not regarded
as independent configuration space variables.


Besides, space-time noncommutativity arising from a relation
like (\ref{10a}), implies that the ``co-ordinate" time $\hat x^0$ cannot
be localised as any state will have a spread in the spectrum of $\hat x^0$.
This leads to the failure of causality and eventually violation of locality
in quantum field theory \cite{balachan}.

\section{Interaction with background Electromagnetic Field}
In this section, we consider interactions with a background electromagnetic
field which still keeps the time reparametrisation symmetry of the relativistic free particle intact. Before discussing the general case, we consider a constant background field. The interaction term to be added to $S_{0}$ is then
\begin{eqnarray}
S_{F} = -\frac{1}{2}\int d\tau F_{\mu\nu}x^{\mu}\dot{x^{\nu}}
\label{42}
\end{eqnarray}
where $F_{\mu\nu}$ is a constant field strength tensor.
The canonical momenta are given by
\begin{eqnarray}
\Pi_{\mu} = p_{\mu} + \frac{1}{2}F_{\mu\nu}x^{\nu}
\label{43}
\end{eqnarray}
where $p_{\mu}$ is given by (\ref{2}).
The reparametrisation symmetry again leads to the Einstein constraint (\ref{4}) which is the first class constraint of the theory. The Poisson brackets are\footnote{These relations follow from the basic canonical algebra $\{x_{\mu}, \Pi^{\nu}\} = \delta_{\mu}^{\nu}; \{x_{\mu}, x_{\nu}\} = \{\Pi_{\mu}, \Pi_{\nu}\} = 0$.}
\begin{eqnarray}
\{x^\mu, p_{\nu}\} = \delta^{\mu}_{\nu} \quad \{x^\mu, x^{\nu}\} = 0 \quad \{p_\mu, p_{\nu}\} = -F_{\mu\nu}
\label{44}
\end{eqnarray}
Note that $p_{\mu}$ does not have zero Poisson bracket with the constraint (\ref{4}) anymore and thus is not gauge invariant. Now to obtain the generator of reparametrisation symmetry, we again exploit the infinitesimal transformation of the space-time coordinate given by (\ref{N8}). Proceeding exactly as in the earlier sections, we write the variation of the Lagrangian in a total derivative form as,
\begin{eqnarray}
\delta L &=& \frac{dB}{d\tau}\qquad ; \qquad B = -m\epsilon\sqrt{-\dot{x}^2} - \frac{\epsilon}{2}F_{\mu\nu}x^{\mu}\frac{dx{^\nu}}{d\tau}
\label{45}
\end{eqnarray}
Then the generator is obtained from usual Noether's prescription (as it was done for the case of the free relativistic particle), by making use of (\ref{43}) to get
\begin{eqnarray}
G = \frac{1}{2}\left(\Pi^{\mu}\delta x_{\mu} - B\right) = \frac{\epsilon\sqrt{-\dot{x}^2}}{2m}\phi_{1}
\label{46}
\end{eqnarray}
where $\phi_{1} = p^2 + m^2 \approx 0$ is the first class constraint (\ref{4}).
Clearly we find that $G$ generates the infinitesimal transformation of the space-time coordinate (\ref{N12a}).
Hence we have again shown that the generator is indeed proportional to the first class constraint which is in agreement with Dirac's treatment. Also, the relation between reparametrisation symmetry and gauge symmetry becomes evident.  
Now the gauge/reparametrisation symmetry can be fixed by imposing a gauge condition. The standard choice is given by (\ref{5}). The constraints (\ref{4}, \ref{5}) form a second class set with the Poisson brackets between them given by (\ref{6}). So the non-vanishing Dirac brackets are given by (\ref{8000}) and
\begin{eqnarray}
\{p_{i}, p_{j}\}_{DB} = -F_{ij}\qquad \{p_{0}, p_{i}\}_{DB} = F_{ij}\frac{p_{j}}{p_{0}}.
\label{46a}
\end{eqnarray}

To obtain non-commutativity between the primed set of space-time coordinates  (\ref{9a}), we first observe that the zeroth component and spatial components of (\ref{N8})(in the standard gauge (\ref{5})) leads to (\ref{9b})
where we have used the relation $\frac{dx^{i}}{d\tau} = -\frac{p_{i}}{p_{0}}$ obtained from (\ref{2}).
Using the relations (\ref{9a}, \ref{9b}) fixes the value of
$\epsilon$, which, in view of the non-vanishing bracket (\ref{46a}), turns out to be
\begin{eqnarray}
\epsilon = -\theta^{0j}P_{j}
\label{47}
\end{eqnarray}
where
\begin{eqnarray}
P_{\mu} = p_{\mu} + F_{\mu\nu}x^{\nu}
\label{48}
\end{eqnarray}
is gauge invariant since $\{P_{\mu}, p_{\nu}\} = 0$. As a simple consistency, observe that for vanishing electromagnetic field, the solution (\ref{47}) reduces to the free particle solution (\ref{9d}). Also note that the non-vanishing Dirac brackets involving $P_{\mu}$ in the standard gauge (\ref{5}) are given by
\begin{eqnarray}
\{x^{i}, P_{j}\}_{DB} = {\delta^{i}}_{j} \quad \{P_{\mu}, P_{\nu}\}_{DB} = F_{\mu\nu}\quad \{x^{i}, P_{0}\}_{DB} = \frac{p^{i}}{p_{0}}
\label{50}
\end{eqnarray}

Using (\ref{47}) we write down the following set of transformations which relate the unprimed and primed coordinates, following from the gauge transformation (\ref{9b}),
\begin{eqnarray}
x^{'0} = x^{0} - \theta^{0i}P_{i}
\label{58}
\end{eqnarray}
\begin{eqnarray}
x^{'i} = x^{i} - \theta^{0j}P_{j}\frac{dx^{i}}{d\tau} = x^{i} + \theta^{0j}P_{j}\frac{p^{i}}{p_{0}}
\label{59}
\end{eqnarray}
where we have used the relation $\frac{p^{'}_{j}}{p^{'}_{0}} = -\frac{dx^{'}_{j}}{d\tau}$ since $\frac{dx^{0}}{d\tau} = 1$ in the old gauge (\ref{5}).
From the above set of transformations and the relations (\ref{8000}, \ref{46a}, \ref{50}), we compute the Dirac brackets between the primed variables
\begin{eqnarray}
\{x^{'0}, x^{'i}\}_{DB} &=& \theta^{0i}
\label{60}
\end{eqnarray}
\begin{eqnarray}
\{x^{'i}, x^{'j}\}_{DB} &=& \frac{1}{p_{0}}\left(\theta^{0i}p^{j} - \theta^{0j}p^{i}\right)\nonumber\\
&=& \frac{1}{p^{'}_{0}}\left(\theta^{0i}p^{'j} - \theta^{0j}p^{'i}\right) + O(\theta^2)
\label{61}
\end{eqnarray}
In order to express the variables on the R.H.S. in terms of primed ones \footnote{Note that, since $P_{\mu}$ (\ref{48}) is gauge invariant, $P^{'}_{\mu} = P_{\mu}$.}, use has been made of (\ref{59}) to get,
\begin{eqnarray}
\frac{p^{'}_{j}}{p^{'}_{0}} = \frac{p_{j}}{p_{0}} - \theta^{0k}P_{k}\frac{d}{d\tau}\left(\frac{p_{j}}{p_{0}}\right) + O(\theta^2)
\label{64a}
\end{eqnarray}

Observe that the change of variables (\ref{58}, \ref{59}) leading to the algebra among the primed variables, are basically infinitesimal gauge transformations that are valid to first order in the reparametrisation parameter $\epsilon$. Moreover, from (\ref{47}) it follows that $\epsilon$ is proportional to $\theta$. Hence, the Dirac algebra (\ref{60}, \ref{61}) between the primed variables are also valid upto order $\theta$. But it turns out that these results are actually exact, as is now shown.

As before, it is possible to write down the modified gauge condition from the solution (\ref{47}) for $\epsilon$ as,
\begin{eqnarray}
\phi_{2} = x^{0} + \theta^{0i}P_{i} - \tau \approx 0, \qquad i = 1, 2, ...d
\label{64aaa}
\end{eqnarray}
The constraints $(\ref{4}, \ref{64aaa})$ again form a second class set with the Poisson brackets between them being given by (\ref{6}). So we recover the previous Dirac brackets (\ref{60}, \ref{61}) between space-time coordinates  $x^\mu$.

Finally we consider the relativistic free particle coupled to an arbitrary electromagnetic field. As before the action is reparametrisation invariant. Here we replace (\ref{42}) by
\begin{eqnarray}
S_{F} = -\int d\tau A_{\mu}(x)\dot{x^{\mu}}
\label{65}
\end{eqnarray}
The choice $A_{\mu} = -\frac{1}{2}F_{\mu\nu}x^{\nu}$ for constant $F_{\mu\nu}$ reproduces the action (\ref{42}).
The Einstein constraint (\ref{4}) and Poisson brackets (\ref{44}) again follow.
The canonical momenta are given by
\begin{eqnarray}
\Pi_{\mu} = p_{\mu} - A_{\mu}
\label{66}
\end{eqnarray}
where $p_{\mu}$ is defined by (\ref{2}). The gauge symmetry can be fixed by imposing a gauge condition. The standard choice is given by (\ref{5}). The constraints (\ref{4}, \ref{5}) form a second class set with the Poisson brackets between them again given by (\ref{6}). So the non-vanishing Dirac brackets are given by (\ref{8000}) and (\ref{46a}).
As before, exploiting the reparametrisation symmetry of the problem, the infinitesimal transformation of the space-time coordinate is given by (\ref{N8}) which leads to (\ref{9b}) in the standard gauge (\ref{5})(where we have again used the relation $\frac{dx^{i}}{d\tau} = -\frac{p^{i}}{p_{0}}$ obtained from (\ref{2})).
Demanding noncommutativity between the primed set of space-time coordinates by imposing the condition (\ref{9a}) and using the relations (\ref{9a}, \ref{9b})
 leads to, 
\begin{eqnarray}
\{x^{0} + \epsilon, x^{i} - \epsilon\frac{p^{i}}{p_{0}}\}_{DB} = \theta^{0i}
\label{6800}
\end{eqnarray}
which fixes the value of
$\epsilon$ to be
\begin{eqnarray}
\epsilon = -\theta^{0j}p_{j} + O(\theta^{2}).
\label{69}
\end{eqnarray}
Here we are content with expression linear in $\theta$ as a gauge invariant $P_{\mu}$ (counterpart of (\ref{48})) cannot be defined here.

Once again we can identify a gauge (which is the same as (\ref{9e}))where we have non-commutativity between space-time coordinates. Computing the Dirac bracket
between the space-time coordinates in this gauge gives,
\begin{eqnarray}
\{x^{0}, x^{i}\}_{DB} = \frac{\theta^{0i}}{1 + \theta^{0j}F_{j\mu}\frac{p^{\mu}}{p_{0}}}
\label{690}
\end{eqnarray}
which has already been given in \cite{pin}.
One can easily see that to the linear order in $\theta$, the above result goes to (\ref{9a}).

\section{Conclusions}
We have discussed an approach whereby both space-space as well as space-time
 noncommutative stuctures are obtained in a particular (nonstandard gauge)
 in models having reparametrisation invariance. These structures are obtained
by calculating either Dirac brackets or symplectic brackets and the results
agree. We have also shown that 
 the noncommutative results in the nonstandard gauge
 and the commutative results in the standard gauge are seen
 to be gauge transforms of each other. In other words, equivalent physics
is described by working either with the usual brackets
 or the noncommuting brackets.
 We feel our approach is conceptually cleaner
 and more elegant than those \cite{pin} where such change of variables are found by inspection and apparently lack any connection with the symmetries
 of the problem. This leads to ambiguities in the definition of physical (gauge invariant) variables. For instance, the angular momentum operator gets modified
 in distinct gauges, by appropriate inclusion of extra terms, so that the closure property is satisfied. In our approach, on the contrary, the angular momentum
 remains invariant since the change of variables
 is just a gauge transformation. Consequently these extra terms never appear.
We feel that the present approach could be useful in illuminating the role
of variable changes used for relating the commuting and noncommuting
descriptions in field theory.

\section*{Appendix A}
Here we would like to demonstrate how the Dirac brackets for any pair of variables, computed for Coulomb and axial gauges, are connected through gauge transformations. For that we consider the action of free Maxwell theory 
\begin{eqnarray}
S = -\frac{1}{4}\int d^{4}x F_{\mu\nu}F^{\mu\nu}
\label{A1}
\end{eqnarray}
Now the first class constraints of the theory are 
\begin{eqnarray}
\pi_{0}(x) \approx 0  \quad \partial_{i}\pi_{i}(x) \approx 0
\label{A2}
\end{eqnarray}
which are responsible for generating gauge transformations. The above set
of constraints can be rendered second class by gauge fixing. Let us first consider the Coulomb gauge which is given by
\begin{eqnarray}
A_{0} \approx 0 \qquad; \qquad\partial_{i}A_{i}(x) \approx 0. 
\label{A3}
\end{eqnarray}
The Dirac bracket computed between $A_{i}, \Pi_{j}$ in this gauge yields the familiar transverse delta function \cite{dirac}, \cite{dir};
\begin{eqnarray}
\{A_{i}(x), \Pi_{j}(y)\}^{(c)}_{DB} &=& -\left(\delta_{ij} - \frac{\partial_{i}\partial_{j}}{\partial^2}\right)\delta (x - y)\nonumber\\
&=& - \delta^{T}_{ij}\delta (x - y) 
\label{A4}
\end{eqnarray}
where the superscript $c$ denotes the Coulomb gauge. 

The corresponding DB in axial gauge $A_{3} \approx 0$ and $(\Pi_{3} - \partial_{3}A_{0}) \approx 0$ \footnote{This follows by demanding time conservation of the gauge; i.e., $\partial_{0}A_{3} = \partial_{0}A_{3} - \partial_{3}A_{0} + \partial_{3}A_{0} = -\Pi_{3} + \partial_{3}A_{0} \approx 0$.}is given by \cite{dir}
\begin{eqnarray}
\{A_{i}(x), \Pi_{j}(y)\}^{(a)}_{DB} &=& -\delta_{ij}\delta (x - y) + \delta_{3j}\frac{\partial_{i}}{\partial_{3}}\delta (x - y)
\label{A8}
\end{eqnarray}

Now the gauge field configurations $A^{(a)}_{i}$ and $A^{(c)}_{i}$ are connected by the gauge transformation 
\begin{eqnarray}
A^{(a)}_{i} = A^{(c)}_{i} + \partial_{i}\Lambda
\label{A5}
\end{eqnarray}
where $\Lambda$ is the gauge transformation parameter.
Imposing $A_{3}^{(a)} = 0$ (axial gauge), fixes the value of $\Lambda$ to be
\begin{eqnarray}
\Lambda = -\frac{1}{\partial_{3}}A^{(c)}_{3}
\label{A6}
\end{eqnarray}
so that,
\begin{eqnarray}
A^{(a)}_{i} = A^{(c)}_{i} - \frac{\partial_{i}}{\partial_{3}}A^{(c)}_{3}
\label{A7}
\end{eqnarray}
On the other hand, $\Pi_{i}$ is gauge invariant, $\Pi_{i}^{(a)} = \Pi_{i}^{(c)}$. Hence, we have,
\begin{eqnarray}
\{A_{i}(x), \Pi_{j}(y)\}_{DB}^{(a)} = \{A_{i}(x)- \frac{\partial_{i}}{\partial_{3}}A_{3}(x), \Pi_{j}(y)\}_{DB}^{(c)}
\label{A70}
\end{eqnarray}
Using the Coulomb gauge result (\ref{A4}), the axial gauge algebra (\ref{A8}) is correctly reproduced.

\section*{Appendix B}
In this appendix, we develop the symplectic formalism and show the 
connection between integral curves and the Hamilton's equations of motion in
the time reparametrised version. 

Let $Q = R \times Q_{0}$, ($Q_{0} = {q^{i}(t), i = 1,2,...,n}$), be a $n + 1$- dimensional configuration space which includes time $t$. The
corresponding phase-space $\Gamma$ is $2n + 2$- dimensional with coordinates
$(t, q^{i}, p_{t}, p_{i})$. On this phase-space, a function $F(t, q^{i}, p_{t}, p_{i})$ is defined as follows
\begin{eqnarray}
F(t, q^{i}, p_{t}, p_{i}) = p_{t} + H_{0}(q^{i}, p_{i})
\label{8}
\end{eqnarray}
Also let $\tilde \theta  = p_{t}dt + p_{i}dq^{i}$ be a $1$-form on $\Gamma$.
Now let $\Sigma$ be a sub-manifold of $\Gamma$ defined by $F(t, q^{i}, p_{t}, p_{i}) = 0$.
Restricting $\tilde \theta$ to $\Sigma$, we get,
\begin{eqnarray}
\tilde \theta|_{\Sigma} = -H_{0}(q^{i}, p_{i})dt + p_{i}dq^{i} 
\label{9}
\end{eqnarray}
An arbitrary tangent vector $\vec{X}$ to a curve in $\Sigma$ is given by
\begin{eqnarray}
\vec{X} = u\frac{\partial}{\partial t} + v^{j}(q^{i}, p_{i})\frac{\partial}{\partial q^{j}} + f_{j}(q^{i}, p_{i})\frac{\partial}{\partial p_{j}}
\label{10}
\end{eqnarray}
with $u$, $v^{j}$ and $f_{j}$'s being arbitrary coefficients.

Demanding that the $2$-form $\tilde \omega = d\tilde\theta|_{\Sigma}$ is degenerate, i.e., $\exists$ $\vec{X} \not= 0$, such that upon contraction, the one-form 
$\tilde \omega (\vec{X}) = 0$, 
we immediately obtain
\begin{eqnarray}
f_{i} + u\frac{\partial H_{0}}{\partial q^{i}} = 0
\label{11}
\end{eqnarray}
\begin{eqnarray}
-v_{i} + u\frac{\partial H_{0}}{\partial p_{i}} = 0
\label{12}
\end{eqnarray}
Hence (\ref{10}) can be written as
\begin{eqnarray}
\vec{X} = u\left(\frac{\partial}{\partial t} + \frac{\partial H_{0}}{\partial p_{i}}\frac{\partial}{\partial q^{i}} - \frac{\partial H_{0}}{\partial q^{i}}\frac{\partial}{\partial p_{i}}\right)
\label{12a}
\end{eqnarray}
Now recall that an integral curve of a vector field is a curve
such that the tangent at any point to this curve gives the value of the vector field at that point.

In general, any tangent vector field $\vec{X}$ to a family of curves,
 parametrised by $\tau$, in the space $\Sigma$ can be 
written as
\begin{eqnarray}
\vec{X} &=& \dot{x}^{\mu}\partial_{\mu} \quad; \quad \dot{x}^{\mu} = \frac{dx^{\mu}}{d\tau}\nonumber\\
&=& \dot{t}\frac{\partial}{\partial t} + \dot{q}^{i}\frac{\partial}{\partial q^{i}} + \dot{p}_{i}\frac{\partial}{\partial p_{i}} 
\label{12b}
\end{eqnarray}
Comparing (\ref{12a}, \ref{12b}), the equations of the integral curves are given by
\begin{eqnarray}
\dot{q}^{i} = u\frac{\partial H_{0}}{\partial p_{i}} \quad;\quad \dot t = u\quad;\quad \dot{p}_{i} = -u\frac{\partial H_{0}}{\partial q^{i}}
\label{12c}
\end{eqnarray}
Note that in the $t = \tau$ gauge, we recover the usual Hamiltonian equations of motion. It is the parameter $u$ which is responsible for inducing the time reparametrisation invariance.

Now we consider the example of a non-relativistic particle in $1 + 1$-dimension, the Hamiltonian of which reads, 
\begin{eqnarray}
H_{0} = \frac{p_{x}^2}{2m}
\label{130}
\end{eqnarray}
In $1 + 1$-dimension, the equations of the integral curves (\ref{12c}) 
can be rewritten as, 
\begin{eqnarray}
\dot{x} = u\frac{\partial H_{0}}{\partial p_{x}}\quad; \quad\dot{t} = u\quad;\quad \dot{p_{x}} = -u\frac{\partial H_{0}}{\partial x}
\label{18}
\end{eqnarray}
Substituting the form of the Hamiltonian (\ref{130}) in (\ref{18}), we obtain,
\begin{eqnarray}
p_{x} = \frac{m\dot x}{\dot t} = m\frac{dx}{dt} = constant
\label{19}
\end{eqnarray}
which is the equation of the integral curve. Note that the above form of the canonical momentum is independent of the parameter $u$. This establishes a connection between the integral curve on $\Sigma$ and the canonical momenta. 
Also from (\ref{8}, \ref{130}), we have,
\begin{eqnarray}
p_{t} + \frac{p_{x}^2}{2m}= 0.
\label{20}
\end{eqnarray}
which is nothing but the first class constraint (\ref{N7}) in the 
time reparametrised version of the non-relativistic particle.
 Hence, from the integral curve, we also get the constraint of the time
reparametrised theory. 
The connection between the integral curves and the constraints
 for the other models discussed in the paper can be
shown in a similar way following the above approach.
\section*{Acknowledgment}

The authors would like to thank the referee for very useful comments.


\end{document}